\documentclass{article}

\PassOptionsToPackage{numbers, compress}{natbib}

\usepackage[preprint]{neurips_2023}
\makeatletter
\renewcommand{\@noticestring}{}
\makeatother

\usepackage[utf8]{inputenc} 
\usepackage[T1]{fontenc}    
\usepackage{hyperref}       
\usepackage{url}            
\usepackage{booktabs}       
\usepackage{amsfonts}       
\usepackage{nicefrac}       
\usepackage{microtype}      
\usepackage{xcolor}         
\usepackage{graphicx}
\usepackage{enumitem}
\title{Multimodal Hybrid Retrieval-Augmented Generation for Scientific Document Understanding using Open-Source SLMs}

\author{%
  Alexandru-Andrei Saucă \\
  Faculty of Electronics, Telecommunications\\
  and Information Technology\\
  University Politehnica of Bucharest\\
  \texttt{alexandru.sauca@stud.etti.upb.ro} \\
  \And
  Ana Luiza Rusnac \\
  Faculty of Electronics, Telecommunications\\
  and Information Technology\\
  University Politehnica of Bucharest\\
  \texttt{ana\_luiza.dumitrescu@upb.ro} \\
}

\begin{document}

\maketitle

\begin{abstract}
   
   Large Language Models tend to hallucinate when answering domain-specific questions from scientific documents without prior fine-tuning. Currently, methods such as Retrieval-Augmented Generation partially solve this problem but face different challenges: limited context knowledge, difference between sparse and dense retrieval, and retrieval noise. This paper presents an Advanced Multimodal Retrieval-Augmented Generation system that aims to solve those challenges and improve the accuracy of information extraction. The proposed architecture introduces a multimodal ingestion pipeline that leverages an open-source Vision-Language Model (Qwen2-VL-2B-Instruct) to generate textual summaries of tables and figures. The retrieval phase integrates HNSW-based semantic search with GIN-based lexical search, unified through Reciprocal Rank Fusion and refined using Cross-Encoder reranking to minimize retrieval noise. To ensure conversational coherence across multi-turn interactions, a Query Condenser module is employed. Evaluation is conducted by independently assessing the ingestion, retrieval and generation stages using the MMLongBench benchmark, a BeIR-format synthetic dataset and the DeepEval framework. Moreover, results demonstrate a 157\% improvement in retrieval quality over a Naive-RAG baseline, with only 50 ms additional latency, while Qwen2-VL-2B-Instruct achieved results comparable to cloud-based models in BERTScore. These findings validate that open-source optimized SLMs, paired with advanced retrieval strategies, can provide competitive performance for document understanding without relying on cloud-based models. 
\end{abstract}

\section{Introduction}

    Generally, most LLM-based applications suffer the same problem, hallucinating when facing domain-specific Q\&A tasks. Furthermore, available documents often contain complex data (texts, tables, figures) that LLMs rarely capture. Solving this problem has driven the adoption of powerful cloud-based models with more resources and capabilities than their open-source competitors, raising costs and introducing privacy concerns. In order to have a reliable, private and affordable system for scientific documents Q\&A, this paper proposes an Advanced Multimodal RAG approach using open-source quantized SLMs for complex contents understanding with a multi-stage hybrid retrieval pipeline. With that idea in mind, the main contributions of the proposed approach are:
    \begin{itemize}[nosep, leftmargin=*, topsep=2pt]
        \item A multimodal data ingestion pipeline leveraging a Vision-Language Model to generate textual descriptions of non-textual elements (tables, figures), enabling unified retrieval across all document content
        \item A hybrid retrieval mechanism combining HNSW-based semantic search with GIN-based lexical search, unified through Reciprocal Rank Fusion with tuned parameters (k=25) and refined using Cross-Encoder reranking
        \item A comprehensive evaluation methodology independently assessing ingestion, retrieval, and generation quality using established benchmarks (MMLongBench, BeIR, DeepEval)
        \item Empirical validation that quantized SLMs achieve competitive performance against cloud-based alternatives for scientific document understanding
    \end{itemize}

\section{Related Work}

    Retrieval-Augmented Generation~\cite{DBLP:journals/corr/abs-2005-11401} 
has been extensively investigated for enhancing NLP tasks by retrieving 
relevant information from a vector store to support generation. However, 
Naive-RAG approaches, relying on single-vector similarity search, are 
prone to hallucinated responses due to limited lexical understanding. 
To address this, Advanced RAG systems employ hybrid retrieval: sparse 
retrieval (BM25)~\cite{BM25, robertson2009probabilistic} for keyword matching and dense retrieval 
(HNSW)~\cite{HNSW} for semantic similarity. Since the scores of these 
methods are incompatible, Reciprocal Rank 
Fusion~\cite{RRF} merges ranked lists regardless of score magnitude, 
while Cross-Encoder reranking~\cite{rerank} provides deeper semantic 
matching to minimize information noise.

Despite these advancements, a significant gap remains in processing 
complex document formats. Traditional RAG systems focus on textual 
ingestion, often discarding non-textual information such as tables and 
figures found in scientific PDFs. While recent studies explore 
Multimodal RAG using Vision-Language Models~\cite{li2025surveystateartlarge}, 
these solutions rely on massive, closed-source API-based models, 
introducing significant limitations regarding data privacy, cost, 
and latency.

To address these gaps, this paper proposes an end-to-end multimodal 
RAG architecture at the intersection of advanced hybrid retrieval and 
efficient multimodal processing. By leveraging locally deployed, 
quantized Small Language Models (SLMs), the proposed system accurately 
processes both textual and visual elements from complex PDFs while 
maintaining strict data privacy and minimizing computational overhead.
    
\section{Approach}

\begin{figure}[htpb]
    \centering
    \includegraphics[width=1.0\textwidth, height=0.2\textheight]{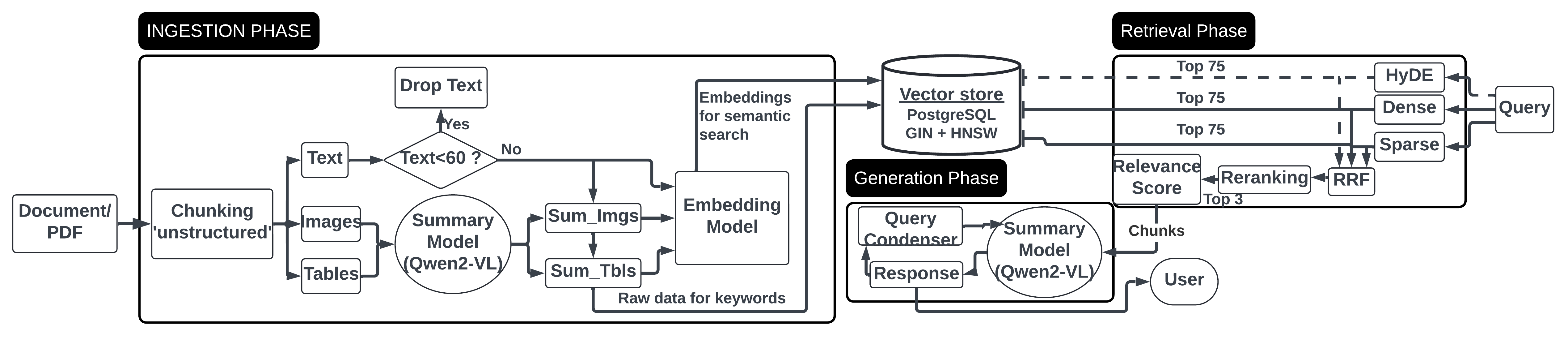}
    \caption{Advanced Multimodal RAG architecture}
    \label{fig:adv_rag}
\end{figure}

    This paper proposes a multimodal RAG architecture composed of four stages: multimodal data ingestion, PostgreSQL vector store, advanced hybrid retrieval pipeline, and response generator. The complete approach operates on locally-deployed, quantized SLMs. Figure \ref{fig:adv_rag} illustrates the complete architecture.

\subsection{Multimodal Data Ingestion}

The ingestion phase addresses the constraints of traditional text-centric RAG pipelines by extracting and vectorizing complex visual elements. As illustrated in Figure \ref{fig:adv_rag}, the process initiates with the parsing of raw PDF documents using an unstructured chunking strategy \cite{unstructured}, categorizing elements into text, images, and tables. To mitigate parsing noise, text chunks are filtered based on a character-length threshold, discarding segments with fewer than 60 characters.

A primary challenge in this stage is the processing of non-textual data, as standard embedding models lack the capability to interpret visual or tabular structures directly. To resolve this, a Vision-Language Model (Qwen2-VL-2B-Instruct \cite{yang2024qwen2technicalreport}) is deployed as a Summary Model to generate descriptive textual representations of the extracted images and tables. Subsequently, these summaries and the valid text chunks are vectorized using EmbeddingGemma-300m \cite{vera2025embeddinggemma}, a 22.7M-parameter embedding model producing 768-dimensional representations. The resulting embeddings are stored in a PostgreSQL database employing a Hierarchical Navigable Small World (HNSW) index to facilitate dense retrieval, while the raw text and summaries are indexed using a Generalized Inverted Index (GIN) for sparse lexical search.

\subsection{Hybrid Retrieval and Reranking}

To overcome the retrieval limitations inherent in single-vector similarity searches, the proposed pipeline utilizes a hybrid methodology that executes parallel queries against the vector store. While the architecture supports query transformation via Hypothetical Document Embeddings (HyDE) \cite{sorstkins2025assessingraghyde1b} —denoted by the dashed pathway in Figure \ref{fig:adv_rag}—the primary operational flow routes the original query directly to the search modules to prioritize low-latency inference for local deployments.

The system performs dense search via the HNSW index to capture semantic similarity and sparse search via the GIN index for exact keyword matching, retrieving the top $k=75$ candidate chunks from each method. Given that semantic distance scores and BM25-like lexical scores operate on disparate scales, Reciprocal Rank Fusion (RRF) with a tuned constant $k{=}25$ is applied to normalize and aggregate the retrieved lists. To maximize Precision@1 and optimize the context window for the generation stage, the aggregated results are processed by a Cross-Encoder Reranker (ms-marco-MiniLM-L6-v2) \cite{CrossEncoderMiniLM}. This model computes a bidirectional relevance score between the query and each candidate chunk, isolating the top 3 highest-scoring contexts.

\subsection{Response Generation}
The final stage handles response synthesis on locally deployed, 
quantized SLMs. To maintain conversational coherence without exceeding 
hardware constraints, a Query 
Condenser~\cite{koo2024optimizingquerygenerationenhanced} reformulates 
the user's latest prompt using prior context into a standalone, fully 
resolved query, avoiding the VRAM overhead and attention dilution of 
appending full conversational history. This condensed query, alongside 
the top 3 retrieved contexts, is supplied to the Generator Model 
(Qwen2-VL-2B-Instruct, INT4 quantized, 1.39~GB VRAM). Through applied 
prompt engineering~\cite{sahoo2025systematicsurveypromptengineering}, 
the generator is constrained to synthesize responses exclusively from 
the provided factual context, promoting high factual fidelity while 
maintaining computational efficiency for consumer-grade deployment.
    
\section{Experimental Analysis}

    \subsection{Experimental Setup}
All experiments were conducted on consumer-grade hardware (Nvidia RTX 
4060 TI, 8~GB VRAM, 32~GB RAM). Three datasets independently assess 
each pipeline stage: MMLongBench~\cite{mmlongbench} 
for VLM summarization quality, a BeIR-formatted synthetic dataset 
generated via the Gemini API for retrieval, and the DeepEval 
framework~\cite{deepeval} for end-to-end generation using an 
LLM-as-a-Judge paradigm. Metrics include ROUGE-LSum, BERTScore F1, 
and LLM-as-a-Judge (1--3 scale) for ingestion; MRR, Recall@5, and 
Precision@1 for retrieval; and Faithfulness, Answer Relevancy, and 
Fluency for generation. Results are benchmarked against a Naive-RAG 
baseline and cloud-based Gemini-2.5-Flash-Lite.\footnote{The codebase and evaluation scripts used in this study are publicly available at: \url{https://github.com/AlexandruSauca/Multimodal-Advanced-RAG-using-local-SLMs-for-PDF-document-understanding}}

\subsection{Multimodal Ingestion Results}
Table~\ref{tab:ingestion} presents the summarization performance of 
various Vision-Language Models on the MMLongBench benchmark \cite{mmlongbench}. The 
selected local model, Qwen2-VL-2B-Instruct \cite{yang2024qwen2technicalreport}, achieves a BERTScore \cite{DBLP:journals/corr/abs-1904-09675} F1 
within 2 points of the cloud-based Gemini-2.5-Flash-Lite baseline 
(55.02 vs.\ 56.96), despite requiring only 1.39~GB of VRAM. While 
Qwen2.5-VL-3B achieves marginally higher BERTScore F1 (56.62), it 
requires 60\% more VRAM (2.23~GB), making Qwen2-VL-2B-Instruct the optimal 
choice for resource-constrained deployment. Furthermore, the results 
reveal a critical discrepancy between n-gram and semantic evaluation 
metrics: Qwen3-2B-Thinking achieves the highest ROUGE-LSum \cite{DBLP:journals/corr/abs-1803-01937} score 
(19.33) but the lowest LLM-Judge rating (1.27), demonstrating the 
inherent limitations of purely lexical metrics for evaluating 
summarization quality.

\begin{table}[htpb]
    \centering
    \caption{VLM Summarization Quality on MMLongBench}
    \label{tab:ingestion}
    \begin{tabular}{@{}lcccc@{}}
        \toprule
        \textbf{Model} & \textbf{VRAM} & \textbf{ROUGE-LSum} & \textbf{BERTScore F1} & \textbf{LLM-Judge} \\ 
        \midrule
        Gemini-2.5-Flash-Lite & Cloud & 14.78 & 56.96 & 2.47 \\
        Qwen2.5-VL-3B & 2.23 GB & 16.77 & 56.62 & 2.02 \\
        \textbf{Qwen2-VL-2B-Instruct (proposed)} & \textbf{1.39 GB} & \textbf{15.06} & \textbf{55.02} & \textbf{1.98} \\
        Gemma3-4B & 3.02 GB & 15.46 & 55.87 & 1.88 \\
        Qwen3-2B-Thinking & 1.46 GB & 19.33 & 52.90 & 1.27 \\ 
        \bottomrule
    \end{tabular}
\end{table}
\vspace{-4pt}

\subsection{Retrieval Evaluation and Ablation Study}
Prior to the architectural ablation study, dense vectorization was optimized by evaluating candidate embedding models. EmbeddingGemma-300m was selected as the primary encoder, outperforming larger alternatives such as Multilingual-E5-Large-Instruct by achieving a superior NDCG@10 of 0.800 and Recall@5 of 0.866, as detailed in Table \ref{tab:embedding_models}. 

Building upon this foundation using the BeIR-formatted synthetic dataset \cite{DBLP:journals/corr/abs-2104-08663}, Table~\ref{tab:retrieval} details the pipeline ablation. The Dense + Sparse hybrid configuration yielded a 157\% MRR improvement over the Naive-RAG baseline (0.132 to 0.340 at Top-K=15). Expanding the initial retrieval pool to Top-K=75 further improved Recall@5 to 0.538; these candidates were normalized via Reciprocal Rank Fusion \cite{RRF} ($k{=}25$) and refined by a Cross-Encoder \cite{CrossEncoderMiniLM} to strictly isolate the top 3 contexts. Notably, at Top-K=75, all configurations---including HyDE-augmented variants---converged to identical performance (MRR: 0.349), rendering HyDE's 30$\times$ latency penalty (3.10s vs.\ 0.10s) unjustifiable for the primary pipeline.

\begin{table}[htpb]
    \centering

    \begin{minipage}[t]{0.38\textwidth}
        \centering
        \caption{Performance of Embedding Models}
        \label{tab:embedding_models}
        \resizebox{\linewidth}{!}{%
        \begin{tabular}{@{}lcc@{}}
            \toprule
            \textbf{Embedding Model} & \textbf{NDCG@10} & \textbf{Recall@5} \\ 
            \midrule
            \textbf{Gemma-300m (proposed)} & \textbf{0.800} & \textbf{0.866} \\
            IBM-Granite-278m & 0.625 & 0.733 \\
            E5-Large-Instruct & 0.704 & 0.769 \\ 
            \bottomrule
        \end{tabular}%
        }
    \end{minipage}\hfill 
    \begin{minipage}[t]{0.58\textwidth}
        \centering
        \caption{Retrieval Pipeline Ablation}
        \label{tab:retrieval}
        \resizebox{\linewidth}{!}{%
        \begin{tabular}{@{}lccccc@{}}
            \toprule
            \textbf{Configuration} & \textbf{Top-K} & \textbf{MRR} & \textbf{Rec@5} & \textbf{Prec@1} & \textbf{Lat.(s)} \\ 
            \midrule
            Naive-RAG (Dense) & 15 & 0.132 & 0.192 & 0.038 & 0.05 \\
            Dense + Sparse & 15 & 0.340 & 0.500 & 0.231 & 0.08 \\
            \textbf{Dense + Sparse (proposed)} & \textbf{75} & \textbf{0.349} & \textbf{0.538} & \textbf{0.231} & \textbf{0.10} \\
            HyDE (Qwen) + Hybrid & 15 & 0.353 & 0.500 & 0.231 & 3.10 \\
            HyDE (Gemini) + Hybrid & 75 & 0.349 & 0.538 & 0.231 & 4.84 \\ 
            \bottomrule
        \end{tabular}%
        }
    \end{minipage}
\end{table}
\vspace{-4pt}

\subsection{Generation Evaluation}
End-to-end generation quality, evaluated via the DeepEval framework \cite{deepeval} (Gemini-2.5-Flash as judge), is detailed in Table~\ref{tab:generation}. The system demonstrates strong factual grounding with an 88.5\% Faithfulness pass rate and 80.8\% Answer Relevancy, confirming the efficacy of the retrieval-augmented pipeline in mitigating hallucinations. While the Fluency pass rate is comparatively lower (69.2\%), this reflects a deliberate architectural trade-off: the highly quantized, 1.39~GB VRAM SLM inherently prioritizes strict factual extraction over stylistic prose, providing an optimal balance for scientific question-answering on resource-constrained consumer hardware.

\begin{table}[htpb]
    \centering
    \caption{Generation Quality Assessment (DeepEval)}
    \label{tab:generation}
    \begin{tabular}{@{}lccc@{}}
        \toprule
        \textbf{LLM} & \textbf{Faithfulness} & \textbf{Answer Relevancy} & \textbf{Fluency}\\ 
        \midrule
        Qwen2-VL-2B-Instruct & 88.5\% & 80.8\% & 69.2\% \\
        \bottomrule
    \end{tabular}
\end{table}
\vspace{-4pt}

\section{Discussion}

    The experimental results reveal several insights regarding the 
viability of locally deployed SLMs for multimodal RAG. The BERTScore 
convergence between Qwen2-VL-2B and cloud-based 
Gemini-2.5-Flash-Lite---despite a difference of two orders of 
magnitude in computational resources---challenges the prevailing 
assumption that competitive document understanding necessarily 
requires large-scale commercial models. This finding is reinforced 
by the retrieval ablation study, where at sufficiently broad retrieval 
depths (Top-K=75), HyDE-augmented configurations yield no measurable 
improvement over standard hybrid retrieval despite introducing 
30$\times$ higher latency, suggesting that retrieval breadth is more 
cost-effective than query transformation for local deployment.

The evaluation methodology itself warrants discussion. The observed 
discrepancy between ROUGE-LSum and LLM-as-a-Judge scores across VLMs 
indicates that n-gram overlap metrics may inadequately capture 
summarization quality in scientific domains, where semantic fidelity 
matters more than lexical similarity. This reinforces recent calls 
for evaluation frameworks that prioritize semantic over surface-level 
assessment~\cite{deepeval}.

Finally, the generation results reveal a consistent 
precision-over-recall pattern: high Faithfulness (88.5\%) but lower 
Fluency (69.2\%). For scientific document question answering, this 
trade-off is arguably desirable---incorrect information carries 
greater risk than stylistically limited responses. However, this 
constrains the system's applicability to domains where linguistic 
polish is secondary to factual accuracy.

\vspace{-6pt}

\section{Conclusion}

   This paper presented an Advanced Multimodal RAG architecture for 
scientific document understanding, demonstrating that locally deployed, 
quantized SLMs can serve as a viable alternative to cloud-based 
solutions. The experimental analysis validates three key findings: 
(1)~hybrid retrieval combining semantic and lexical search achieves a 
157\% improvement over Naive-RAG baselines, (2)~the quantized 
Qwen2-VL-2B model attains summarization quality within 2 BERTScore 
points of commercial APIs while requiring only 1.39~GB VRAM, and 
(3)~broader retrieval pools effectively subsume the benefits of query 
expansion techniques such as HyDE, eliminating unnecessary latency. 
These results suggest that effective retrieval engineering can 
compensate for reduced model scale, offering a practical path toward 
private, low-cost scientific document understanding. Future 
investigations include evaluation on larger domain-diverse benchmarks, 
migration to more capable open-source VLMs, and containerized 
deployment for production scalability.

\section{Limitations}

    This work presents several limitations that should be considered 
when interpreting the results. First, regarding \textbf{dataset 
constraints}, the retrieval evaluation relies on a synthetically 
generated BeIR-formatted dataset produced via the Gemini API, 
which may not fully capture the complexity and diversity of 
real-world scientific queries. Similarly, the generation 
evaluation comprises a limited number of test cases, reducing 
the statistical power of the reported pass rates. The 
MMLongBench benchmark focuses predominantly on government 
reports, and the system's performance on other scientific 
domains (e.g., biomedical, legal) remains unvalidated.

Second, regarding \textbf{method constraints}, the multimodal 
parsing pipeline occasionally fails to preserve the spatial 
relationship between non-textual elements (tables, figures) and 
their surrounding textual context, leading to summaries that 
lack critical contextual information. Furthermore, the limited 
context window of the quantized SLM restricts the amount of 
retrieved content that can be processed simultaneously, which 
may result in information loss for queries requiring synthesis 
across multiple document sections.

Third, regarding \textbf{methodological trade-offs}, the system 
has been evaluated exclusively using automated metrics 
(BERTScore, DeepEval, ROUGE); no human evaluation was conducted 
to validate perceived response quality. Additionally, the 
current architecture has been tested on a single consumer-grade 
GPU and has not been evaluated under production-scale 
concurrent workloads, leaving scalability characteristics 
unexplored.
    
\section{AI Transparency Statement}

Generative AI was utilized across the methodology, evaluation, and manuscript preparation. The proposed system employs Qwen2-VL-2B-Instruct for multimodal summarization and generation. For evaluation, the Gemini-2.5-Flash API was used to generate synthetic retrieval datasets and function as an LLM-as-a-judge via DeepEval. Finally, Google Gemini 3.1 Pro and Claude AI Opus 4.6 assisted in manuscript preparation strictly for structural refinement, code generation, and iterative feedback on section drafts. All technical content, experimental results, and architectural decisions are the original work of the author.
    
\begin{ack}
This work was conducted as part of a Bachelor's thesis at 
University Politehnica of Bucharest. The author would like 
to thank Ana Luiza Rusnac for her supervision and guidance 
throughout this project.
\end{ack}

\small{
\bibliography{references}
\bibliographystyle{plainnat}
}

\end{document}